\documentclass[conference]{IEEEtran}
\IEEEoverridecommandlockouts

\usepackage{cite}
\usepackage{amsmath,amssymb,amsfonts}
\usepackage{algorithmic}
\usepackage{graphicx}
\usepackage{subcaption}
\usepackage{caption}
\usepackage{textcomp}
\usepackage{xcolor}
\usepackage{url}
\usepackage{siunitx}
\usepackage{comment}

\def\BibTeX{{\rm B\kern-.05em{\sc i\kern-.025em b}\kern-.08em
    T\kern-.1667em\lower.7ex\hbox{E}\kern-.125emX}}

\begin{document}

\title{AI-Enabled Smart Hygiene System for Real-Time Glucose Detection}

\author{\IEEEauthorblockN{Khan Masood Parvez}
\IEEEauthorblockA{\textit{Electronics  \& }\\
\textit{Communication Engineering} \\
\textit{Aliah University}\\
Kolkata, India \\
masoodrph@gmail.com}
\and
\IEEEauthorblockN{Sk Md Abidar Rahaman}
\IEEEauthorblockA{\textit{Computer Science \&}\\
\textit{Engineering} \\
\textit{University of Burdwan}\\
West Bengal, India \\
abidar.ce@gmail.com}
\and
\IEEEauthorblockN{Ali Shiri Sichani}
\IEEEauthorblockA{\textit{Electrical Engineering \&}\\ \textit{Computer Science} \\
\textit{University of Missouri}\\
Columbia, USA \\
asp9f@missouri.edu}
\and
\IEEEauthorblockN{Hadi AliAkbarpour}
\IEEEauthorblockA{\textit{Computer Science} \\
\textit{Saint Louis University}\\
Saint Louis, USA \\
hadi.akbarpour@slu.edu}
}

\maketitle

\begin{abstract}
This research presents a smart urinary health monitoring system incorporating a coplanar waveguide (CPW)-fed slot-loop antenna biosensor designed to analyse various urine samples. The antenna demonstrates distinct resonant frequency shifts when exposed to five specific urine conditions, deviating from its baseline 1.42 GHz operation. These measurable frequency variations enable the antenna to function as an effective microwave sensor for urinary biomarker detection. A potential artificial intelligence-based Convolutional Neural Networks - Long Short-Term Memory (CNN-LSTM) framework is also discussed to overcome the limitations of overlapping frequency responses, aiming to improve the accuracy of health condition detection. These components contribute to the development of a smart toilet system that displays real-time health information on a wall-mounted urinal screen, without requiring any user effort or behavioural change.
\end{abstract}

\renewcommand\IEEEkeywordsname{Keywords}
\section{Introduction}
The integration of artificial intelligence (AI) with microwave sensing technologies is poised to revolutionize personalized healthcare by enabling non-invasive, real-time monitoring of chronic conditions like diabetes mellitus (DM). With over 589 million global cases reported [1], the need for frequent glycemic tracking is critical to prevent complications such as neuropathy, retinopathy, and cardiovascular disease. Traditional methods like fingerstick tests and continuous glucose monitors (CGMs) remain invasive, costly, and uncomfortable. To address these issues, this study presents a smart toilet system for real-time urinary glucose analysis using microwave sensor and machine learning (ML) algorithms, offering a non-invasive, hygienic, and continuous monitoring solution. Microwave sensors have demonstrated promise in biomedical applications, with advancements in high-Q resonators and differential designs improving sensitivity [2]–[5]. Integrated, low-cost systems support real-time, non-invasive analysis [3], [6]–[9]. Innovations such as SRRs, SIW, and biomimetic materials further enhance resolution and compactness [10]–[15]. CPW-fed slot-loop antennas and dielectric resonator antennas (DRAs) have been optimized for performance, including improved bandwidth and radiation patterns [16]–[22]. AI-based medical sensors continue to advance diagnostics and patient monitoring [23]–[25], and wearable biosensors offer non-invasive glucose detection for diabetes care [26]. Smart toilets are emerging as a practical health monitoring tool, integrating AI, biosensors, and imaging to track physiological and disease markers [27]–[30].

The rest of the paper is organized as follows. Section II describes the design of a coplanar waveguide-fed slot-loop antenna for sensitive glucose monitoring.  Section III presents an AI-assisted approach for discriminating overlapping frequency signatures. Section IV introduces the concept and implementation of the smart toilet system. Finally, Section V concludes the paper and highlights future research directions.
\begin{figure}[!t]
  \centering
  \includegraphics[width=\linewidth]{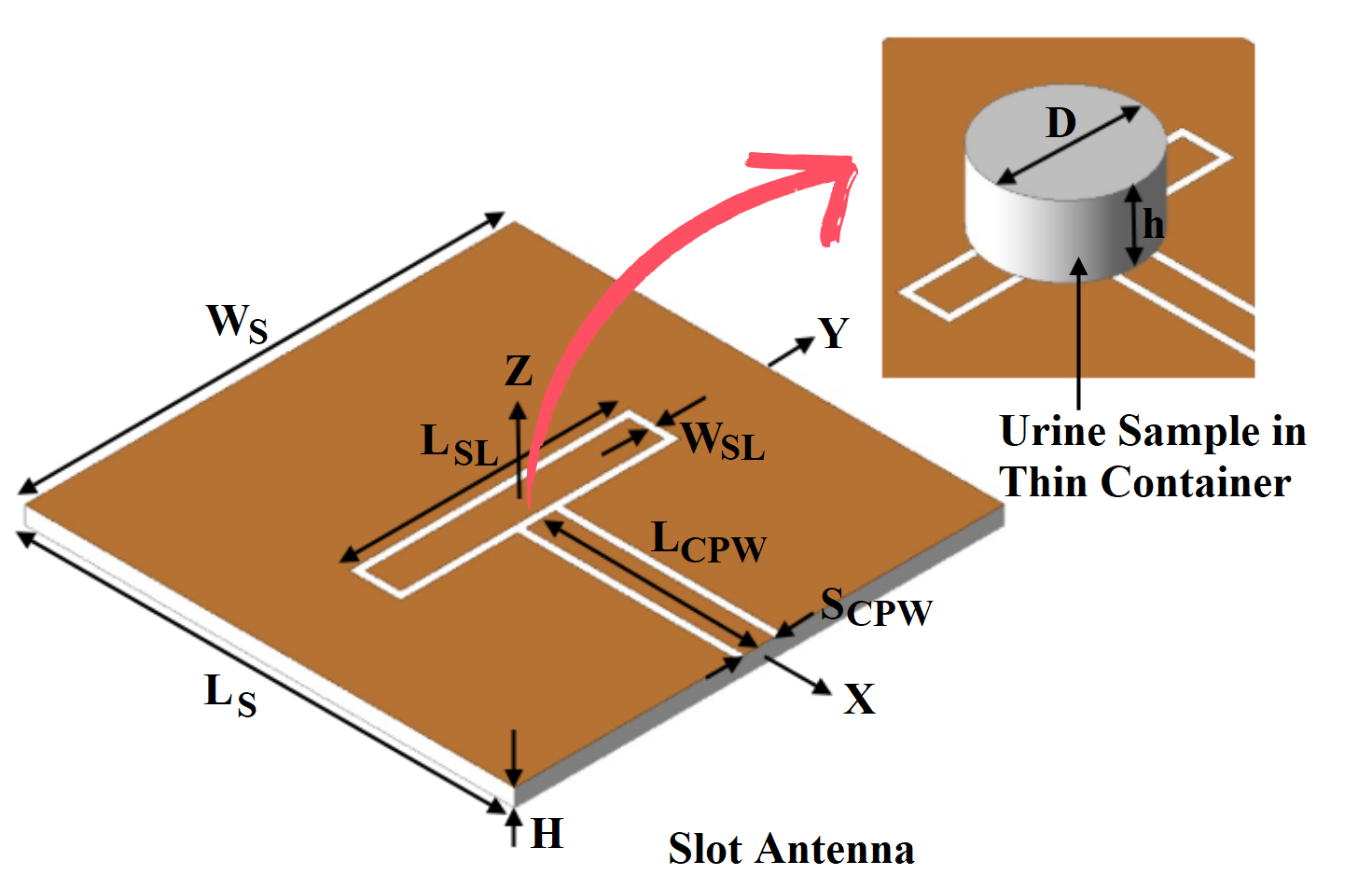} 
  \caption{Design of the coplanar waveguide-fed slot-loop antenna and intrigated with urin sample}
  \label{fig:antenna_design}
\end{figure}
\begin{table}[h]
    \centering
    \caption{Dielectric Properties of Urine at $1 \, \text{GHz}$   and \ 25$^{\circ}$C.} 
    \label{tab:urine_properties}
    \begin{tabular}{@{}lll@{}}
        \hline 
        \textbf{Condition} & \textbf{Parameter} & \textbf{Value} \\
        \hline 
        Healthy (Normal)         & Permittivity       & $70$--$75$ \\
                                 & Conductivity       & $0.5$--$2.0$ Siemens/m \\
                                 & Density            & $1.010$--$1.030$ g/cm³  \\
        \hline
        Diluted (Low Density)    & Permittivity       & $75$--$80$ \\
                                 & Conductivity       & $0.2$--$0.5$ Siemens/m \\
                                 & Density            & $1.000$--$1.010$ g/cm³ \\
        \hline
        Concentrated (High Density) & Permittivity    & $60$--$70$ \\
                                    & Conductivity    & $2.0$--$4.0$ Siemens/m \\
                                    & Density         & $1.025$--$1.040$ g/cm³ \\
        \hline
        Diabetic (Glucose-Rich)  & Permittivity       & $65$--$72$ \\
                                 & Conductivity       & $1.5$--$3.5$ Siemens/m \\
                                 & Density            & $1.020$--$1.035$ g/cm³\\
        \hline
        Dehydrated (High Salt)   & Permittivity       & $55$--$65$ \\
                                 & Conductivity       & $3.0$--$5.0$ Siemens/m \\
                                 & Density            & $1.030$--$1.050$ g/cm³ \\
        \hline 
    \end{tabular}
\end{table}
\begin{figure}[!t]
  \centering
  \includegraphics[width=0.6\linewidth]{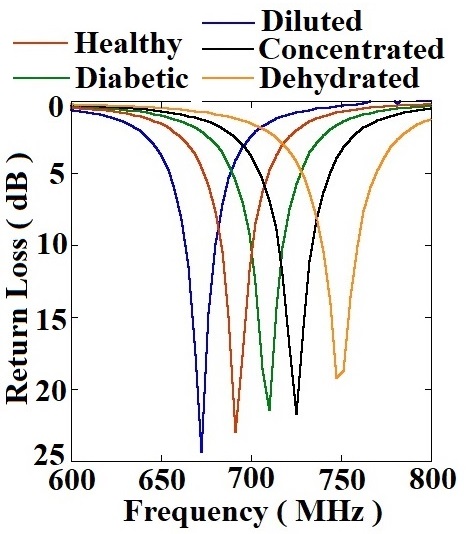}
  \caption{Return loss plot for different permittivity values corresponding to varying health conditions of urine samples}
  \label{fig:antenna_design}
\end{figure}
\section{Coplanar Waveguide-Fed Slot-Loop Antenna for Sensitive Glucose Monitoring }
 Fig. 1 illustrates the geometry of the Coplanar Waveguide-Fed Slot-Loop Antenna [18], which integrates a CPW feed with a slot-loop resonator for efficient radiation. The antenna is fabricated on a Duroid 6010LM substrate (\(\varepsilon_r = 10.2\), \(H = {2.54} \) mm), chosen for its low-loss properties, making it suitable for high-frequency operation with minimal signal loss. The antenna designs were simulated in Ansys High-Frequency Structure Simulator (HFSS), incorporating a finite copper conductivity of \(5.8 \times 10^7\) Siemens/m for the ground plane and a substrate dielectric loss tangent (\(\tan\delta\)) of 0.0023.
 The dimensions are imillimetresrs:
$\mathit{L_S} = \mathit{W_S} = {70} $,
$\mathit{L_{SL}} = {40} $,
$\mathit{W_{SL}} = {1.2} $,
$\mathit{L_{CPW}} = {31.3}$,
and $\mathit{S_{CPW}} ={4.2}$. The antenna employs a coplanar waveguide (CPW) feedline to directly excite a square slot-loop etched into the ground plane. Both the feedline and radiating elements are integrated on the same plane, reducing profile and simplifying fabrication. The design ensures efficient energy coupling and precise resonant frequency.
\begin{table*}[ht]
    \centering
    \caption{Urine Conditions, Associated Diseases/States, and Key Clinical Indicators}
    \label{tab:urine_states}
    \renewcommand{\arraystretch}{1.2} 
    \begin{tabular}{|p{3.8cm}|p{6cm}|p{6.4cm}|}
        \hline
        \textbf{Urine Condition} & \textbf{Associated Diseases/States} & \textbf{Key Clinical Indicators} \\
        \hline
        Healthy Human Urine &
        No disease (baseline) &
        Normal hydration, balanced electrolytes\\
        \hline
        Diluted Urine & Overhydration, diabetes insipidus (ADH deficiency) &
        Low specific gravity (\(<1.010\)), pale color, high urine  \\
        \hline
        Concentrated Urine & Dehydration, fever/sweating, high-protein diet, SIADH (syndrome of inappropriate ADH) &
        Dark color, high specific gravity (\(>1.025\)), low urine volume \\
        \hline
        Diabetic Urine & Diabetes mellitus (Type 1/2), gestational diabetes &
        Glycosuria, hyperglycemia, high urine osmolality \\
        \hline
        Dehydrated Urine & Severe dehydration, acute kidney injury (pre-renal), high salt intake, Addison’s disease (low aldosterone) &
        Elevated Na\textsuperscript{+}/K\textsuperscript{+}, high urine osmolality, low urine volume \\
        \hline
    \end{tabular}
\end{table*}
The CPW-fed slot-loop antenna exhibits a fundamental resonance at 1.42 GHz. The resonant frequency \(f_r\) of a slot antenna is given by:

\begin{equation}
f_r = \frac{c}{2 L_{\text{eff}} \sqrt{\varepsilon_{\text{eff}}}}
\label{eq:slot_resonant_frequency}
\end{equation} where \(c\) is the speed of light in free space, \(L_{\text{eff}}\) is the effective slot length, and \(\varepsilon_{\text{eff}}\) is the effective dielectric constant of the substrate.
 Table 1 lists the dielectric properties—permittivity, conductivity, and density—of human urine at 25°C and 1 GHz, measured across five physiological conditions. The average permittivity values obtained were utilized in the CPW-fed slot-loop antenna design. When exposed to urine samples under these five conditions, the antenna exhibits distinct resonance frequency shifts, beginning from 1.42 GHz, enabling non-invasive health monitoring. Specifically, healthy urine resonates at 691.25 MHz, diluted urine at 672.5 MHz, diabetic at 710 MHz, dehydrated urine at 747.5 MHz and concentrated urine at 725 MHz. Fig. 2 confirms these variations through the return loss plot, indicating strong suitability for microwave sensing. Table 2 summarizes the correlation between urine conditions, related medical disorders, and key clinical indicators. Accurate assessment of glucose levels in urine relies on the permittivity of the sample, as it significantly influences the resonant frequency of the CPW-fed slot-loop antenna, enabling non-invasive detection. It is a well-known fact that the parallel-plate capacitor technique offers a reliable means to determine the complex permittivity \(\epsilon^* = \epsilon' - j\epsilon''\) of biological fluids such as urine.
\author{\IEEEauthorblockN{Your Name}
\IEEEauthorblockA{Your Affiliation\\
Email: your.email@example.com}}
\maketitle
\section{Potential AI-Assisted Discrimination of Overlapping Frequency Signatures}
A major challenge in non-invasive glucose detection is distinguishing diabetic urine from dehydrated urine, as both produce similar frequency shifts (\( \Delta f \approx 35\text{--}40\,\mathrm{MHz} \)). However, their dielectric origins differ: diabetic urine causes a significant \( S_{11} \) phase shift  \(15^\circ\) due to dipole relaxation, while dehydration leads to a smaller shift (less than \(5^\circ\))  driven by ionic changes. Relying solely on frequency shift (\( \Delta f \)) risks false positives, especially since most commercial VNAs disregard sub-1\,GHz phase data, missing key biomarkers.

Glucose primarily affects the dielectric loss component (\( \varepsilon'' \)), whereas dehydration impacts the storage component (\( \varepsilon' \)). To address this spectral overlap, a phase-aware deep learning framework is proposed, using raw complex \( S_{11} \) data—both magnitude and phase—as input to a CNN-LSTM model. This approach extracts features such as phase rotation (\( \Delta \phi \)) and Q-factor variation, improving discrimination between similar responses.
\begin{figure}[htbp]
    \centering

    \textbf{(a) Accuracy and Loss Vs Epoch of CNN-LSTM } \\[4pt]
    \includegraphics[width=0.9\linewidth]{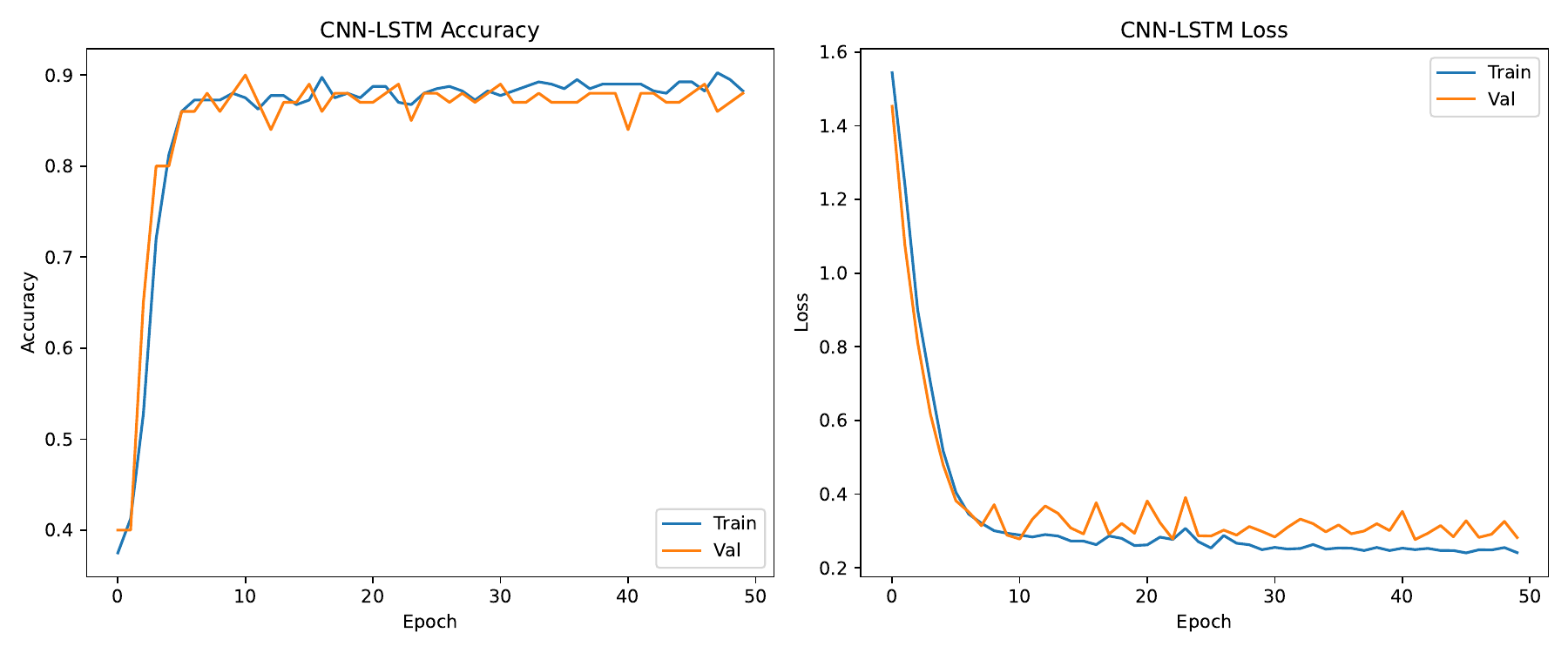}
    \vspace{10pt}

    \textbf{(b) Confusion Matrix and Classification report of CNN-LSTM} \\[4pt]
    \begin{subfigure}[t]{0.48\linewidth}
        \centering
        \includegraphics[width=\linewidth]{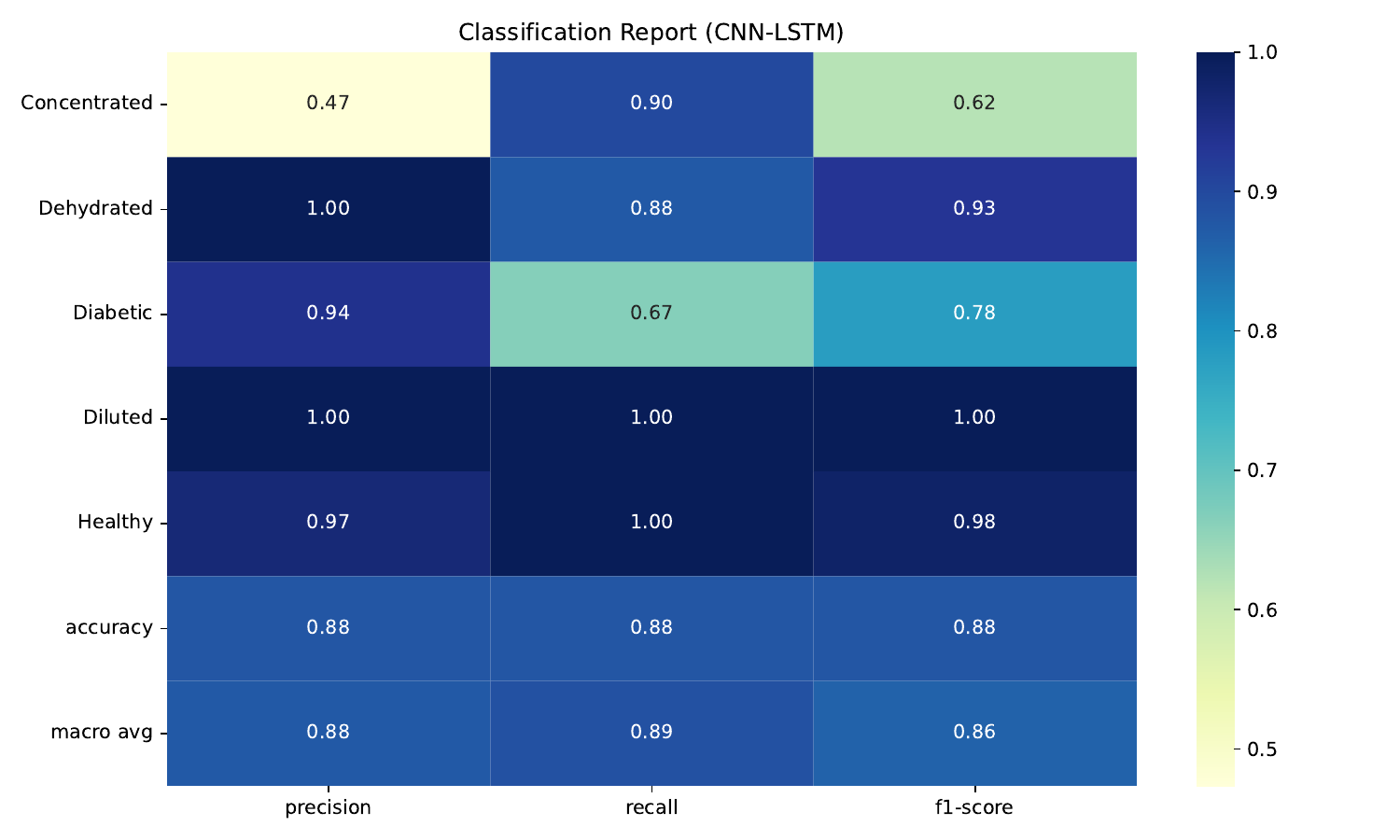}
        \caption{Classification Report}
    \end{subfigure}
    \hfill
    \begin{subfigure}[t]{0.48\linewidth}
        \centering
        \includegraphics[width=\linewidth]{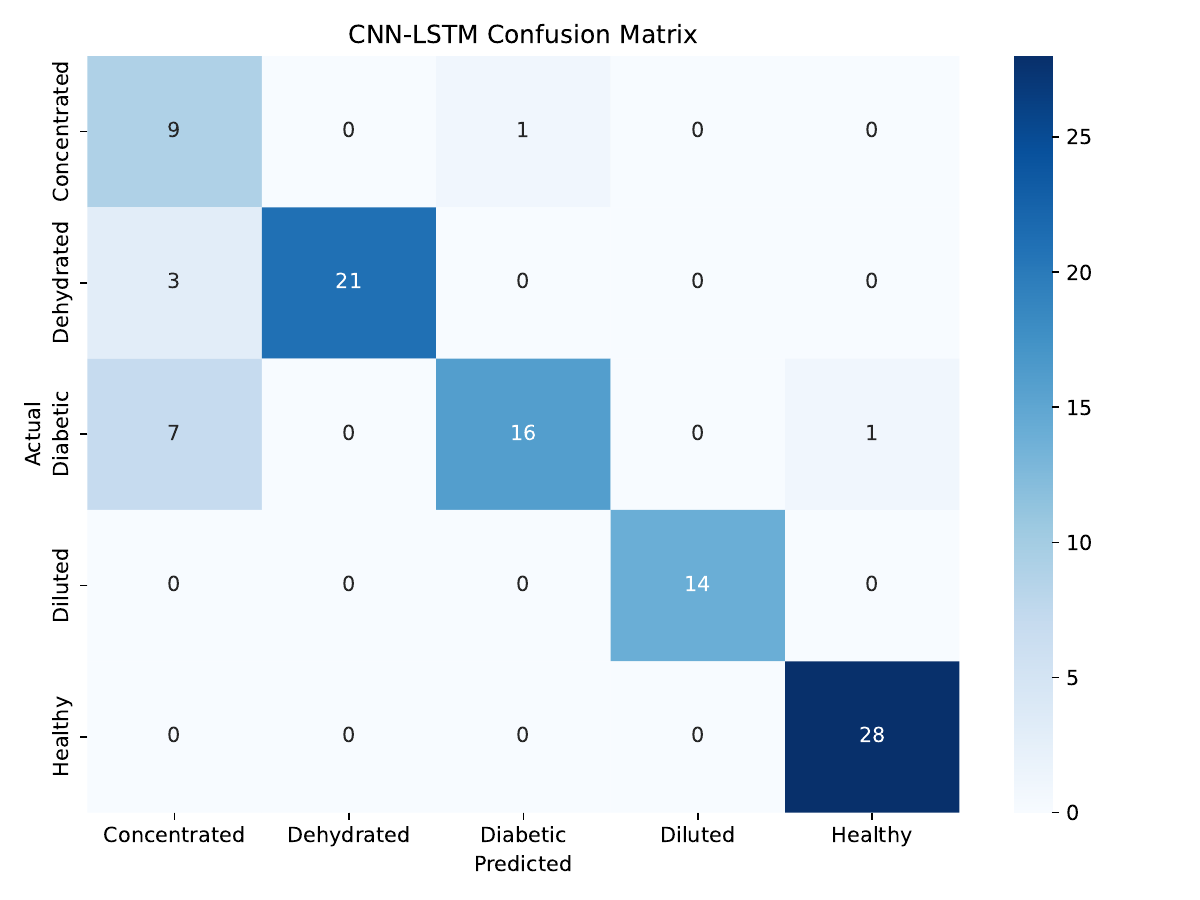}
        \caption{Confusion Matrix}
    \end{subfigure}

    \caption{Different efficiency metrics of CNN-LSTM}
\end{figure}

The CNN-LSTM model demonstrates promising performance across various metrics. The accuracy of the model increases steadily during training, with training accuracy reaching approximately 0.9 and validation accuracy closely following this trend, indicating effective learning. Training loss decreases from around 1.6 to below 0.4, while validation loss shows a similar improvement, reflecting good generalization. The confusion matrix reveals that the model performs excellently on the ``Healthy'' and ``Diluted'' classes, with no misclassifications in ``Healthy'' and only a single misclassification in ``Diluted.'' However, there are some misclassifications in the ``Concentrated'' and ``Diabetic'' classes, where the model struggles with precision and recall. The classification report shows that the ``Diluted''and ``Healthy'' classes achieve perfect precision 1.0 and recall 1.0, with F1-scores of 1.00, indicating a good balance. The ``Concentrated'' class has a precision of 0.47, and the ``Diabetic''class has a recall of 0.67, suggesting areas for improvement in these categories. The overall model accuracy is 0.88, indicating solid performance but with potential room for further optimization in the ``Concentrated'' and  ``Diabetic'' classes. Overall, the model demonstrates strong capabilities, particularly in predicting the ``Healthy'' and ``Diluted'' classes, while additional adjustments could enhance its performance on other classes. Fig.3 presents various efficiency metrics of the CNN-LSTM model. 
\begin{table}[htbp]
\centering
\caption{CNN-LSTM Classification Metrics}
\renewcommand{\arraystretch}{1.2}
\begin{tabular}{|l|c|}
\hline
\textbf{Metric}           & \textbf{Score} \\
\hline
Accuracy                 & 0.88         \\
Precision (Macro)        & 0.88         \\
Recall (Macro)           & 0.89         \\
F1-Score (Macro)         & 0.86        \\
\hline
\end{tabular}\label{tab:cnn_lstm_metrics}
\end{table}
Table \ref{tab:cnn_lstm_metrics} summarizes the performance of the proposed CNN-LSTM model on the classification task. The model achieved an accuracy of 0.88, with macro-averaged precision and recall scores of 0.88 and 0.89, respectively, and a macro F1-score of 0.86. These metrics indicate robust and reliable performance across all classes.

\section{Smart Toilet }
The innovative smart urinary health monitoring platform revolutionises monitoring by embedding advanced biosensors within standard wall-mounted large urinal pot facilities. During routine use, these non-invasive sensors perform real-time urinalysis, with results instantly displayed via digital interfaces. This system effectively addresses the limitations of traditional glucose monitoring methods – eliminating the discomfort of fingerstick blood tests and the expense of CGM systems while maintaining clinical-grade accuracy. The technology enables passive metabolic monitoring during routine restroom use, requiring no user effort or behaviour change.
\section{Conclusion}
This paper introduced a smart  urinary health monitoring system featuring a CPW-fed slot-loop antenna and a potential AI-based framework for real-time, non-invasive urine glucose detection. The antenna exhibited frequency responses sensitive to glucose levels, and urine permittivity was analyzed accordingly. Although the envisioned hybrid CNN-LSTM model could enhance classification using return loss ( $S_{11}$ ) data, its implementation was limited by the lack of a large, controlled dataset. Future work will focus on clinical data collection and advanced AI models, such as transformer-based architectures, to improve the system’s accuracy and scalability for metabolic health monitoring.

\end{document}